\documentclass[12pt]{iopart}

\usepackage{graphicx}
\usepackage{here}
\usepackage{stackrel}
\usepackage{amssymb,accents,latexsym}

\newcommand*{\dt}[1]{%
  \accentset{\mbox{\bfseries .}}{#1}}
\newcommand*{\ddt}[1]{%
  \accentset{\mbox{\bfseries .\hspace{-0.25ex}.}}{#1}}

\newcommand{\diag}{{\textrm{diag}}}

\newcommand{\TeV}{\textrm{\scriptsize TeV}}
\newcommand{\Planck}{\textrm{\scriptsize Planck}}

\newcommand{\h}{\mathfrak h}
\newcommand{\SM}{{\textrm{\scriptsize SM}}}
\newcommand{\ind}{{\textrm{\scriptsize ind}}}
\newcommand{\hol}{{\textrm{\scriptsize hol}}}
\newcommand{\UV}{{\textrm{\scriptsize UV}}}
\newcommand{\IR}{{\textrm{\scriptsize IR}}}

\newcommand{\LD}{{\textrm{\scriptsize LD}}}
\newcommand{\EFT}{{\textrm{\scriptsize EFT}}}
\begin{document}

\title[On gapped continuum resonance spectra]{On gapped continuum resonance spectra}

\author{Eugenio~Meg\'{\i}as$^{1}$ and Mariano~Quir\'os$^{2}$}
\address{
$^{1}$ Departamento de F\'{\i}sica At\'omica, Molecular y Nuclear and Instituto Carlos I de F\'{\i}sica Te\'orica y Computacional, Universidad de Granada, \\
Avenida de Fuente Nueva s/n, 18071 Granada, Spain \\

$^{2}$ Institut de F\'{\i}sica d'Altes Energies (IFAE) and Barcelona Institute of  Science and Technology (BIST), Campus UAB, 08193 Bellaterra, Barcelona, Spain
}
\ead{emegias@ugr.es, quiros@ifae.es}

\vspace{10pt}
%\begin{indented}
%\item[]August 2017
%\end{indented}

\begin{abstract}
In this work we study a warped five-dimensional (5D) model with ultraviolet (UV) and infrared (IR) branes, that solves the hierarchy problem with a fundamental 5D Planck scale $M$, and curvature parameter $k$, of the order of the 4D Planck mass $M_{\textrm{\scriptsize Planck}}\equiv 2.4\times 10^{15}$ TeV. The model exhibits a continuum of Kaluza-Klein (KK) modes with different mass gaps, at the TeV scale, for all fields. We have computed the Green's functions and spectral densities, and shown how the presence of a continuum KK spectrum can produce an enhancement in the cross section of some Standard Model processes. The metric is linear near the IR, in conformal coordinates, as in the linear dilaton and 5D clockwork models, for which $M\sim$ TeV. We also analyze a pure (continuum) linear dilaton scenario, solving the hierarchy problem with more conventional  fundamental $M$ and $k$ scales of the order of $M_{\textrm{\scriptsize Planck}}$, and a continuum spectrum.
\end{abstract}

%
% Uncomment for keywords
%\vspace{2pc}
\noindent{\it Keywords}: physics beyond the Standard Model, extra dimensions, AdS/CFT, Higgs boson, spectral functions 
%
% Uncomment for Submitted to journal title message
%\submitto{\JPA}
%
% Uncomment if a separate title page is required
%\maketitle
% 
% For two-column output uncomment the next line and choose [10pt] rather than [12pt] in the \documentclass declaration
%\ioptwocol
%

\section{Introduction}
\label{sec:introduction}

The Standard Model (SM) of particle physics fails to describe a number of theoretical and experimental issues, such as the existence of dark matter and the strong sensitivity of the electroweak (EW) scale to ultraviolet (UV) physics, i.e.~the EW hierarchy problem. Thus, there is theoretical and experimental evidence to believe that the SM is not the ultimate theory, but an effective theory which works at scales below a few TeV. The hierarchy problem has motivated the study of several extensions of the SM, the most popular ones being supersymmetry and theories with a warped extra dimension. In the Randall-Sundrum (RS) model~\cite{Randall:1999ee}, the hierarchy between the Planck and the EW scale is generated by a warped extra dimension in Anti de Sitter (AdS) space, and it is conjectured to be dual (by the AdS/CFT correspondence) to conformal 4D theories, with composite Higgs boson and towers of composite resonances, i.e.~the so-called Kaluza-Klein (KK) modes.

Direct searches of new physics have traditionally focused on the detection of bumps in the invariant mass of final states. However the elusiveness of isolated and narrow heavy resonances at the LHC~\cite{Aaboud:2019roo} has led people to study different solutions. An exploring possibility to cope with this situation is the existence of a (quasi)-continuum of KK states: this is the idea behind the clockwork models~\cite{Giudice:2017suc,Giudice:2017fmj} and the linear dilaton (LD) models~\cite{Antoniadis:2011qw,Csaki:2018kxb}, which predict an (almost) continuum spectrum with a TeV mass gap. If the new physics consists in a continuum of states, its presence should be associated, not with a bump but with an excess, with respect to the SM prediction, in the measured cross sections. The larger the mass gap, the higher energy should one produce to detect this excess. 
In these models the 5D Planck scale $M$, as well as the 5D curvature constant $k$ are fixed (not so conventionally) at the TeV, while the 4D Planck scale, $M_{\Planck}$, is a derived one with an accidentally large value triggered by a large warping factor enhancing the TeV scales. 

In this work we pursue a more conventional approach and propose a model where $M\simeq k\simeq M_{\Planck}$ are fundamental scales, and the TeV gap is obtained after warping down the scale $k$. The back reaction of the scalar field on the metric generates a linear dilaton only in the IR region, while in the UV the behavior is AdS$_5$. This permits a holographic interpretation of the model and connections with unparticles~\cite{Georgi:2007si,Cacciapaglia:2007jq,Falkowski:2009uy}. Finally, let us mention that in our theory the Higgs boson has an isolated narrow resonance, and a continuum of states above a TeV mass gap, so that it can be considered as a modelization of theories, dubbed Unhiggs theories, which share those features~\cite{Stancato:2008mp,Falkowski:2008yr}. Finally, we will also consider a 5D theory where the dilaton is linear in conformal coordinates for all the interval, solving in a distinctive way the hierarchy problem,  but keeping the behavior $M\simeq k\simeq M_{\Planck}$. We will also point out the limitations of this theory.

\section{Warped extra-dimensional models}
\label{sec:Randall_Sundrum}

In this section we will first provide a short introduction to the RS model, and  propose the warped model that will be used in the rest of this work. We also give some details about the Higgs sector and the EW symmetry breaking in the model.

\subsection{The Randall-Sundrum model}
\label{subsec:Randall_Sundrum}

The RS model was proposed as a theory that allows to solve the hierarchy problem in particle physics~\cite{Randall:1999ee}. The model is based on a 5D space with line element
\begin{equation}
ds^2 = e^{-2A(y)} \eta_{\mu\nu} dx^\mu dx^\nu - dy^2 \,, \qquad A(y) = ky \,,
\end{equation}
and two branes, each of them located at a different position in the extra dimension~$y$. The UV/IR brane location $y=y_0/y_1$ corresponds to the Planck/TeV scale, respectively, and the hierarchy between these scales is driven by the warped space geometry, i.e.
\begin{equation}
M_{\TeV} = e^{-k y_1} M_{\Planck} \,.
\end{equation}
In order to solve the hierarchy problem, the brane dynamics should fix $k (y_1-y_0) \simeq 35$, and this implies $M_{\TeV} \simeq 10^{-15} M_{\Planck}$. The degree of compositeness/elementariness of the fields depends on their location in the extra dimension. Fields that are mainly localized toward the IR brane are composite: this is the case of the Higgs, heavy fermions and KK modes. On the contrary, fields that are localized toward the UV brane are elementary, as e.g. light fermions. Other fields, like the zero mode gauge bosons, are flat in the extra dimension.

One of the main features of the RS model is that the brane distance is stabilized by the so-called Goldberger-Wise mechanism, in which a bulk scalar field $\phi$ breaks conformal invariance with bulk and brane potentials fixing its vacuum expectation value~\cite{Goldberger:1999uk}. A consequence is the appearance of a {\it ``light state''}: the radion/dilaton, with an interesting Higgs-like phenomenology, see e.g.~\cite{Csaki:2000zn,Megias:2014iwa,Megias:2015ory}.

In spite of these interesting properties, the naive RS model has some phenomenological problems as, for instance, it fails to describe oblique observables $S,T,U$, which are related to electroweak precision measurements. A possible way out that we will pursue in this work is to consider large back reaction on the metric such as to create a singularity~\cite{Cabrer:2011fb}, leading e.g.~to a warp factor of the form
\begin{equation}
A(y) \propto \log\left[1 - \frac{y}{y_s} \right]  \,. 
\end{equation}
The singularity of the metric at $y = y_s $ is admissible in the sense of Ref.~\cite{Gubser:2000nd}, as it supports finite temperature in the form of a black hole horizon.

\subsection{The gravitational background}
\label{subsec:grav_back}

Let us introduce the model that we will use in this work. We consider a scalar-gravity system with two branes at values $y = y_0$ (UV brane), and $y = y_1$ (IR brane)~\footnote{As we will explain below, in addition the metric will have an admissible singularity placed at $y = y_s$, such that $y_0 < y_1 < y_s$.} where we are conventionally fixing $y_0=0$. The 5D action of the model reads~\cite{Cabrer:2009we}
\begin{eqnarray}
\hspace{-2cm} S_\phi &=& \int d^5x \sqrt{|\det g_{MN}|} \left[ -\frac{1}{2\kappa^2} R + \frac{1}{2} g^{MN}(\partial_M \phi)(\partial_N \phi) - V(\phi) \right] \nonumber \\
\hspace{-2cm} &&\hspace{2cm}- \sum_{\alpha} \int_{B_\alpha} d^4x \sqrt{|\det \bar g_{\mu\nu}|} \lambda_\alpha(\phi)  
 -\frac{1}{\kappa^2} \sum_{\alpha} \int_{B_\alpha} d^4x \sqrt{|\det \bar g_{\mu\nu}|} K_\alpha    \,.
\end{eqnarray}
There are three kind of contributions to the action, corresponding to the bulk, the brane and the Gibbons-Hawking-York contribution. $V(\phi)$ is the bulk scalar potential, $\lambda_\alpha(\phi)$ $(\alpha=0,1)$ are the UV and IR 4D brane potentials located at $\phi_\alpha\equiv \phi(y_\alpha)$, and $\kappa^2 = 1/(2M^3)$ with $M$ being the 5D Planck scale. The metric is defined as
\begin{equation}
ds^2 = g_{MN}dx^M dx^N \equiv  e^{-2A(y)} \eta_{\mu\nu} dx^\mu dx^\nu-dy^2 \,, \label{eq:metric}
\end{equation}
where the 4D induced metric is $\bar g_{\mu\nu} = e^{-2A(y)} \eta_{\mu\nu}$, and the Minkowski metric is given by $\eta_{\mu\nu} = \diag(1,-1,-1,-1)$. In terms of the metric of Eq.~(\ref{eq:metric}), the extrinsic curvature reads as $K_{0,1} = \mp 4 A^\prime(y_{0,1})$.

The classical equations of motion (EoM) in the bulk can be written in terms of the superpotential, $W(\phi)$, leading to two first order equations~\cite{DeWolfe:1999cp}
\begin{equation}
\phi^\prime(y) = \frac{1}{2} \frac{\partial W}{\partial \phi} \,, \qquad A^\prime(y) = \frac{\kappa^2}{6} W \,,  \label{eq:phiA}
\end{equation}
and
\begin{equation}
V(\phi) = \frac{1}{8} \left( \frac{\partial W}{\partial \phi} \right)^2 - \frac{\kappa^2}{6} W^2(\phi) \,.  \label{eq:V}
\end{equation}
We will consider the model in the stiff wall limit, which means that $\lambda_\alpha(\phi)$, chosen to fix the brane minima at values $\phi(y_\alpha) = v_\alpha$, are given by $\lambda_\alpha(\phi)=\gamma_\alpha(\phi-v_\alpha)^2$ in the limit $\gamma_\alpha\to\infty$.
In addition, we will assume a $\mathbb Z_2$ symmetry $(y \to -y)$ across the UV brane, which translates into the following boundary conditions of the fields
\begin{equation}
A^\prime(y_0) = \frac{\kappa^2}{6} \lambda_0(\phi_0) \,, \qquad \phi^\prime(y_0) = \frac{1}{2} \frac{\partial\lambda_0(\phi_0)}{\partial\phi} \,,
\end{equation}
while the localized term in the IR brane imposes the following jumping conditions
\begin{equation}
\Delta A^\prime(y_1) = \frac{\kappa^2}{3} \lambda_1(\phi_1) \,,  \qquad \Delta \phi^\prime(y_1)  = \frac{\partial\lambda_1(\phi_1)}{\partial\phi} \,,
\end{equation}
where $\Delta X(y_1) \equiv X(y_1^+) - X(y_1^-)$ is the function jump.

\subsection{The soft-wall model}
\label{subsec:soft_wall}

We will use this formalism for a kind of soft-wall phenomenological models defined by the superpotential
\begin{equation}
W(\phi) = \frac{6k}{\kappa^2} \left(1 +  e^{\nu \phi} \right) \,, \label{eq:W_softwall}
\end{equation}
where $\nu$ is a real parameter. The EoM of Eq.~(\ref{eq:phiA}) can be solved analytically with this ansatz, leading to the following profile for the scalar field and warp factors
\begin{equation}
\phi(y) = -\frac{1}{\nu} \log\left[ \frac{3k\nu^2}{\kappa^2} (y_s - y) \right] \,, \qquad A(y) = ky - \frac{\kappa^2}{3\nu^2} \log\left[1 - \frac{y}{y_s} \right]  \,, 
\end{equation}
where we have chosen $A(0) = 0$. Note that the solution of the hierarchy problem demands that $A_1 \equiv  A(y_1) \simeq 35$, and near the UV boundary, i.e.~for $y \ll y_s$, the geometry is AdS$_5$: $A(y)\simeq k y$. The IR singularity of $A(y)$ at $y =y_s$ was already anticipated in Sec.~\ref{subsec:Randall_Sundrum}. The structure of the model in the extra dimension can be summarized as follows
\begin{eqnarray}
&&y=0 \quad \hspace{1.0cm} < \hspace{1.5cm} y_1  \hspace{1.5cm} < \hspace{1.5cm} y_s   \nonumber \\
&& \hspace{-0.6cm} (\textrm{UV brane})  \hspace{2.2cm}  (\textrm{IR brane})  \hspace{1.5cm}  (\textrm{IR singularity})  \nonumber 
\end{eqnarray}
The parameter~$\nu$ is responsible for the properties of the spectrum of the KK modes of the fields. More in details, one can distinguish between three different situations: i)~$\nu > \kappa/\sqrt{3}$ which leads to discrete KK spectra with TeV spacing; ii)~$\nu < \kappa/\sqrt{3}$ corresponding to ungapped continuum KK spectra similar to unparticles; and iii)~$\nu = \kappa/\sqrt{3}$, which is a critical case corresponding to continuum KK spectra with a gap. In the following we will focus on the latest case. 

When considering conformally flat coordinates, i.e.~$dy = e^{-A(z)} dz$, in the critical case the scalar field and warp factor behave near the IR singularity $(z \to +\infty)$ as
\begin{equation}
\phi(z) \simeq \frac{\sqrt{3}}{\kappa} \rho z \,, \qquad A(z) \simeq \rho z \,,
\end{equation} 
where $\rho \equiv e^{-ky_s}/y_s$ is a scale of the order of the TeV. In this way, both $\phi$ and $A$ behave linearly in terms of the conformal coordinate in the IR region, a common property with the linear dilaton models, see e.g.~Refs.~\cite{Cox:2012ee,Giudice:2016yja,Giudice:2017fmj}.

\subsection{The Higgs sector}
\label{subsec:Higgs_sector}

The Higgs sector can be introduced in the theory by defining the 5D bulk doublet for the Higgs~\cite{Cabrer:2011fb}
\begin{equation}
H(x,y)=\frac{1}{\sqrt 2}e^{i \chi(x,y)} \left(\begin{array}{c}0\\h(y)+ \widehat{H}(x,y)
\end{array}\right)  \,,
\end{equation} 
where $h(y)$ is the background Higgs field, $\widehat{H}(x,y)$ is the Higgs excitation and the matrix $\chi$ contains three SM Goldstone fields. The action of the model is $S = S_\phi + S_H$, with
\begin{equation}
\hspace{-2.2cm} S_H = \int d^5x\sqrt{|\det g_{MN}|} \left[ |D_M H|^2 - V(H) \right]-\int d^4x\sqrt{-g_{\ind}}(-1)^\alpha \lambda^\alpha(H)\delta(y-y_\alpha) \,, \label{eq:S_Higgs}
\end{equation}
where~$V(H)= M^2(\phi)|H|^2$ is the 5D Higgs potential, and electroweak symmetry breaking is triggered by the brane potentials~$\lambda^0(H)= 2 M_0|H|^2$ and $\lambda^1(H)= M_1|H|^2-\gamma|H|^4$.

By considering a convenient choice of the bulk potential, one finds the following profile for the Higgs background, $h(y) \simeq \sqrt{\frac{M_1}{\gamma}}  e^{a k (y - y_1)}$, where $a>2$ is a real parameter, which is located toward the~IR. The next step is to impose the correct electroweak symmetry breaking, and this demands
\begin{equation}
v^2=\int_0^{y_s} dy \, h^2 e^{-2 A}  \qquad \textrm{with} \qquad v = 246 \, \textrm{GeV} \,.
\end{equation}
Moreover, by constructing the effective 4D theory, one finds the SM Higgs potential~\cite{Cabrer:2011fb}
\begin{equation}
\hspace{-1.5cm} V_{\SM}=-\mu^2|H_{\SM}|^2+\lambda |H_{\SM}|^4 \,, \qquad \textrm{where} \qquad H(x,y) \sim \frac{h(y)}{h(y_s)} \frac{k}{\rho} H_{\SM}(x) \,,
\end{equation}
and therefore one can relate the Higgs mass, $m_H = 125 \,\textrm{GeV}$, with the model parameters. This leads  to $m_H \simeq \sqrt{M_1/k} \cdot \rho$, from where one can see that the ``natural'' value of $m_H$ would be $\rho$, unless one tunes the mass $M_1/k$ to values $\sim 0.01 - 0.1$, providing this parameter a measure of the size of the ``unnaturalness'' in this class of theories. Finally, the oblique $S$ and $T$ parameters can be computed along the lines of Ref.~\cite{Cabrer:2011fb}. The experimental bounds turn out to be well reproduced at $95\%$ C.L. for $a > 2.3$ and $\rho > 1.25 - 2.5 \,\textrm{TeV}$, depending on the particular value of $a$~\cite{Megias:2019vdb}.

\subsection{The continuum linear dilaton model}
\label{subsec:linear_dilaton}

While the model of Sec.~\ref{subsec:soft_wall} typically leads to expressions for the Green's functions and other observables that demand a numerical treatment, it is possible to consider a simplified version of the model that allows for an analytical study. This is the case of the superpotential and bulk potential given by
\begin{equation}
W(\phi) =  \frac{6k}{\kappa^2} e^{\nu \phi}=\frac{6k}{\kappa^2} e^{\bar \phi},\qquad V(\phi)=-\frac{9k^2}{2\kappa^2}e^{2\bar \phi} \,, \label{eq:W_lineardilaton}
\end{equation}
where we are considering $\nu=\kappa/\sqrt{3}$ and define $\bar\phi\equiv \kappa \phi/\sqrt{3}$, which are the same as the ones for the soft-wall model, cf.~Eq.~(\ref{eq:W_softwall}), but neglecting the cosmological constant term. 

This model looks similar to the linear dilaton model of Refs.~\cite{Cox:2012ee,Giudice:2016yja} except that for the latter the coefficient is $\nu<0$, and therefore the 4D Planck scale is a derived quantity from the 5D Planck scale and the scale $k$, which are supposed to be of order the TeV. Moreover the spectrum in the linear dilaton model is discrete. 

However, the model defined by superpotential (\ref{eq:W_lineardilaton}) has a singularity at a finite value of the proper coordinate $y=y_s$, as in soft wall models. It leads to a gapped continuum spectrum, and the hierarchy problem is, more conventionally, solved in the same way as in RS theories, with fundamental scales $M$ and $k$ of the order of $M_{\Planck}$, and a derived TeV scale after warping. The solution for the background in proper coordinates is
\begin{equation}
\bar\phi(y)=-\log[k(y_s-y)],\quad A(y)=-\log(1-y/y_s) \,.
\end{equation}
We will also consider, as in general soft wall models, two branes at $y=0$ and $y=y_1$, such that the values of the IR brane location $y_1$ and the singularity $y_s$ will be dynamically determined by brane potentials ·$\lambda_\alpha(\phi)$, fixing the field $\bar\phi$ at the values $\bar v_0$ and $\bar v_1$ in the UV and IR branes, respectively, such that
\begin{equation}
ky_s=e^{-\bar v_0},\quad ky_1=e^{-\bar v_0}-e^{-\bar v_1} \,.
\end{equation}
Moreover, the solution of the hierarchy problem is achieved for $A(y_1)\equiv A_1\simeq 35$, which imposes the relation
\begin{equation}
\bar v_1 - \bar v_0 = A_1 \,,
\end{equation}
while the gap of the continuum is given by $\rho=1/y_s=k\,e^{-A_1}$ for $\bar v_0\simeq -A_1$ and $\bar v_1\simeq 0$.

In conformally flat coordinates the background is given by~\footnote{The relation between conformal and proper coordinates in the model of Eq.~(\ref{eq:W_lineardilaton}) turns out to be $z - z_0 = -y_s \log( 1 - y/y_s)$.}
\begin{equation}
\bar\phi(z) =  \rho \cdot (z-z_0)  - \log\left(  k/\rho \right)  \,, \qquad A(z) = \rho \cdot (z - z_0) \,, \label{eq:phiALD}
\end{equation} 
where we fix $z_0=1/k$ and $\rho$ is at the TeV scale. In conformally flat coordinates the location of the branes in units of $\rho$ are at $\rho z_0=e^{-A_1}$ and $\rho z_1 = A_1 + e^{-A_1}\simeq A_1$,~\footnote{Note the difference with respect to the RS geometry for which $\rho z_1=1$.} while the singularity is located at the infinite $z_s\to\infty$. The location of the IR brane is dynamically fixed, in conformal coordinates, by $\bar v_0$ and $\bar v_1$ as $kz_1=1+(\bar v_1-\bar v_0)e^{-\bar v_0}$.
In Sec.~\ref{sec:Linear_Dilaton_Green} of this work we will obtain analytical results for Green's functions in this model.

\section{Holographic Green's functions}
\label{sec:Green_functions}

Let us consider the scalar part of the propagator of a particle with isolated poles as
\begin{equation}
\frac{1}{p^2 - m^2 + i \epsilon} = \mathcal{P} \frac{1}{p^2 - m^2} + i\pi \delta(p^2 - m^2) \,. \label{eq:propagator}
\end{equation}
In this expression $\mathcal{P}$ denotes principal value. The Green's functions that we will study in the following generalize Eq.~(\ref{eq:propagator}) to propagators with an isolated pole (the zero mode) and a continuum of states (instead of a discrete sum of KK modes) with a mass gap~$m_g$, i.e.
\begin{equation}
G(p^2,m_g^2) = \textrm{Re} \, G(p^2,m_g^2) + i \, \textrm{Im} \, G(p^2,m_g^2) \, \theta(p^2 - m_g^2) \,. \label{eq:G_general}
\end{equation}
The gap $m_g \sim \,\textrm{TeV}$ is linked to the solution of the hierarchy problem.~\footnote{Eq.~(\ref{eq:G_general}) is precisely the behavior of gapped unparticles.} We will compute in this section the Green's function for the different kinds of fields, in the model introduced in Secs.~\ref{subsec:soft_wall} and \ref{subsec:Higgs_sector}, by considering the holographic method. We will separately analyze the cases of KK gauge bosons, fermions, graviton, radion and the Higgs boson.

\subsection{Gauge bosons}
\label{subsec:gauge_bosons}

\begin{figure*}[htb]
 \includegraphics[width=0.43\textwidth]{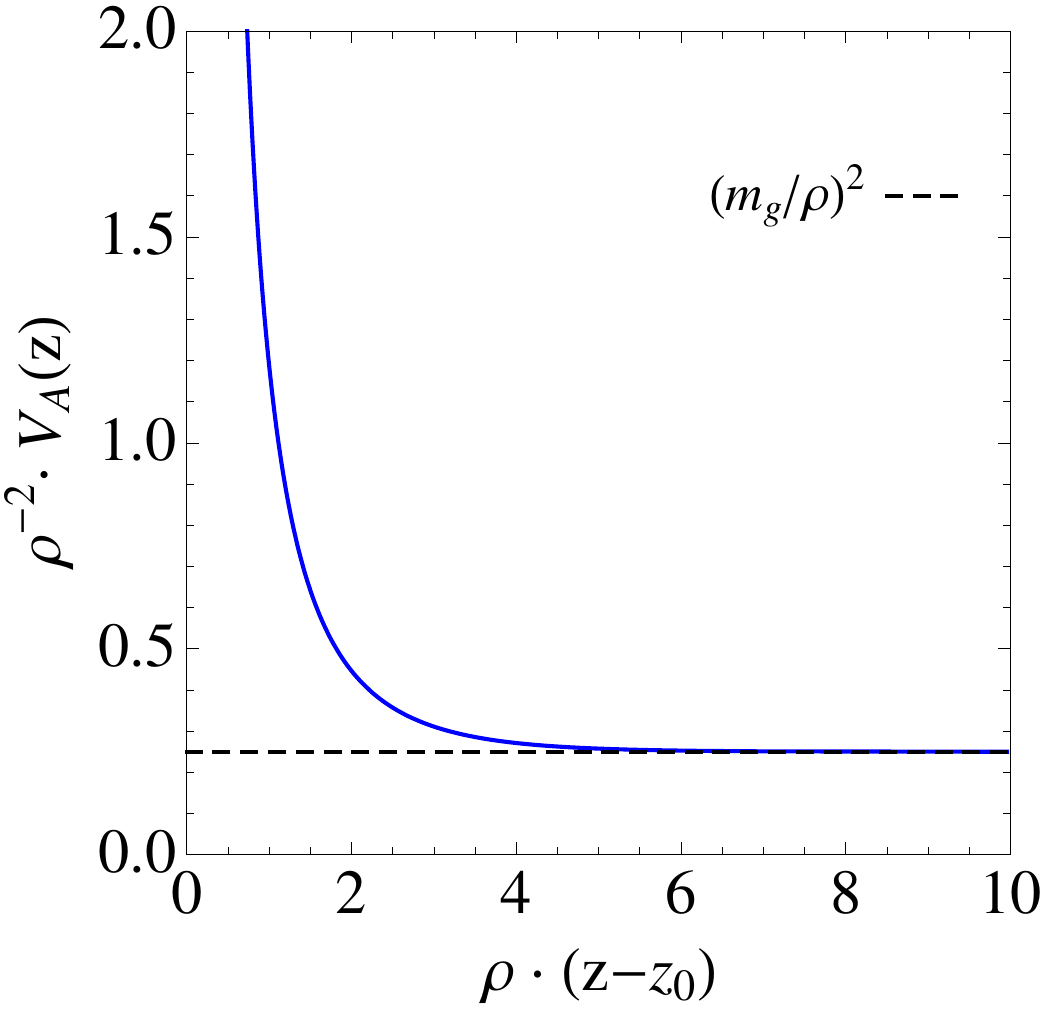} 
\hspace{-1cm} \begin{minipage}{23pc}
\vspace{-3.3cm} \caption{\it Effective Schr\"odinger potential for massless gauge bosons $V_A$ as a function of $z$ as given by Eq.~(\ref{eq:VA}). }
\label{fig:VA}
\end{minipage}
\end{figure*}

Let us consider the Lagrangian for massless gauge bosons (i.e.~the SM photon and gluon), which is given by
\begin{equation}
\mathcal L= \int_0^{y_s} dy\left[ -\frac{1}{4}F_{\mu\nu}F^{\mu\nu}-\frac{1}{2}e^{-2A}A'_\mu A'_\mu  \right]\,.
\end{equation}
Defining $A_\mu(x,y)=f_A(y) A_\mu(x)$, in conformal coordinates and after an appropriate rescaling of the field, one obtains the Schr\"odinger like form for the EoM of the fluctuations 
\begin{equation}
-\ddt{f}_A(z) + V_A(z) f_A(z) = p^2 f_A(z) \,, \label{eq:EOM_fA}
\end{equation}
where the dot denotes derivative with respect to $z$, and potential 
\begin{equation}
V_A(z) =  \frac{1}{4} \dt{A}(z)^2 - \frac{1}{2} \ddt{A}(z)  \,, \qquad  V_A(z) \stackrel[z \to \infty]{\longrightarrow}{} \left( \frac{\rho}{2} \right)^2 \,. \label{eq:VA}
\end{equation}
The behavior of this potential is displayed in Fig.~\ref{fig:VA}. Note that $V_A(z)$ is bounded from below by the value indicated in Eq.~(\ref{eq:VA}), and the corresponding spectrum is a continuum of states above the mass gap $m_g =\rho/2$. 

To compute the holographic Green's function, let us consider the fluctuations in momentum space
\begin{equation}
A_\mu(p,z)=f_A(p,z)a_\mu^{(4)}(p) \,.
\end{equation}
If we fix the boundary condition at the UV brane as~$A_\mu(p,z_0) \equiv a_\mu^0(p)$, where $a_\mu^0$ is a source coupled to the CFT vector operator~$\mathcal J_\mu^A$, the holographic Lagrangian turns out to be
\begin{equation}
\mathcal L_{\hol}=\frac{1}{2}\frac{\dt{f_A}(p,z_0)}{f_A(p,z_0)}\,a_\mu^{0}P^{\mu\nu}(\xi)a_\nu^{0} \,,
\end{equation}
where~$P^{\mu\nu}(\xi)=\eta^{\mu\nu}-(1-1/\xi)p^\mu p^\nu/p^2$ and $\xi$ is a gauge parameter. Finally, the two-point function is the inverse of this bilinear operator, i.e.
\begin{equation}
G^{\mu\nu}_A(z_0,z_0;p)=[\eta^{\mu\nu}-(1-\xi)p^\mu p^\nu/p^2]G_A(z_0,z_0;p) \,,
\end{equation}
where the 4D Green's function and spectral density are
\begin{equation}
G_A(z_0,z_0;p)=\frac{f_A(p,z_0)}{\dt{f_A}(p,z_0)} \,, \qquad  \rho_A(z_0,z_0;p)=\frac{1}{\pi}\, \textrm{Im} \, G_A(z_0,z_0;p) \,. 
\end{equation}
One can define as well a scale invariant Green's function, i.e.~invariant with respect to variation of the parameter $\rho$, as $\mathcal G_A(z_0,z_0;p) \equiv  (\rho^2/k) \mathcal W(k/\rho) G_A (z_0,z_0;p)$ where $\mathcal W(z)$ is the Lambert function. 

\begin{figure*}[htb]
\centering
 \begin{tabular}{c@{\hspace{3.5em}}c}
 \includegraphics[width=0.43\textwidth]{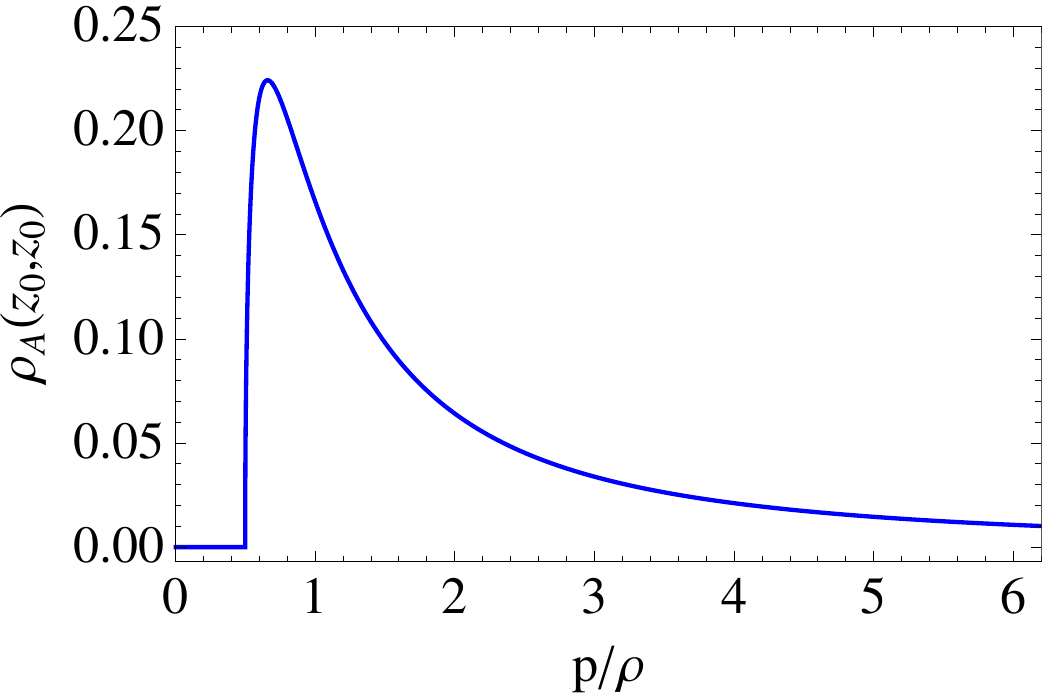} &
\includegraphics[width=0.43\textwidth]{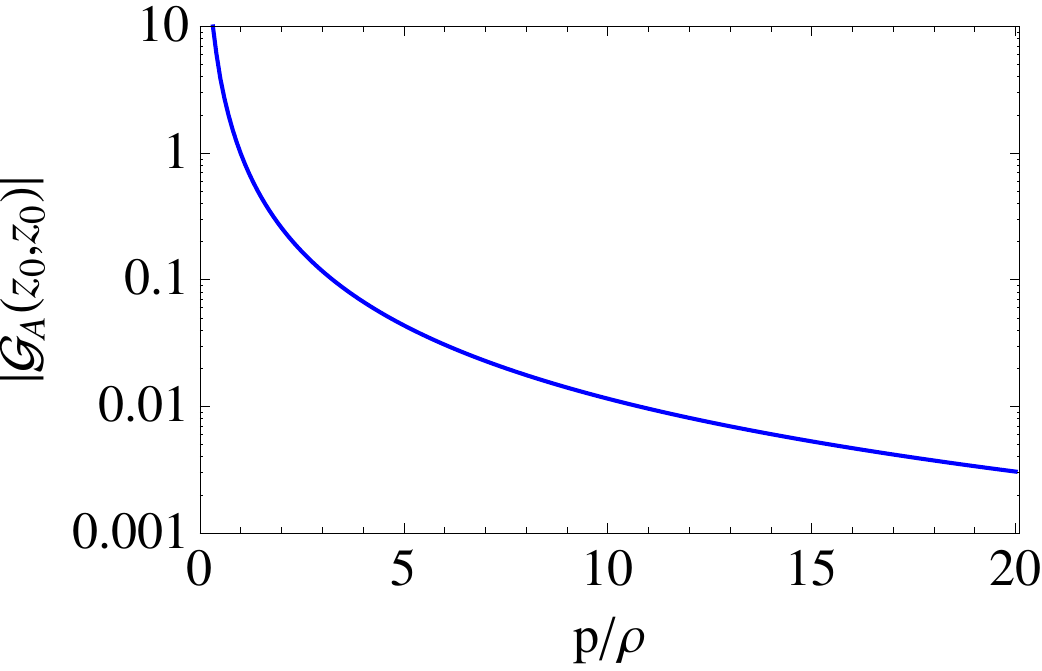} 
\end{tabular}
 \caption{\it Scale invariant spectral density $\rho_A(z_0,z_0;p)$ (left panel) and absolute value of the Green's function $\mathcal G_A(z_0,z_0;p)$ (right panel) for a continuum massless gauge boson.}
\label{fig:spectral_gaugebosons}
\end{figure*}

The solution of Eq.~(\ref{eq:EOM_fA}) in the IR is of the form
\begin{equation}
f_A(z) \simeq  c_- e^{-\frac{1}{2}\Delta \rho  z} + c_+ e^{\frac{1}{2}\Delta \rho z} \,, \quad  \textrm{ with } \quad  \Delta = \sqrt{1- (2p/\rho)^2}\,.\label{eq:fA_IR}
\end{equation}
The computation of the retarded Green's function demands the use of ``IR regular'' solutions for Euclidean AdS, i.e.~$c_+=0$, and this corresponds to outgoing wave boundary conditions after analytical continuation~\cite{Son:2002sd}. Using this we have solved numerically Eq.~(\ref{eq:EOM_fA}), and the results for the scale invariant spectral density and Green's function for massless gauge bosons are shown in Fig.~\ref{fig:spectral_gaugebosons}. Note that the existence of the mass gap is manifest in the spectral density, as it vanishes for $p < \rho/2$.

\subsection{Fermions}
\label{subsec:fermions}

\begin{figure*}[htb]
\centering
 \begin{tabular}{c@{\hspace{3.5em}}c}
 \includegraphics[width=0.43\textwidth]{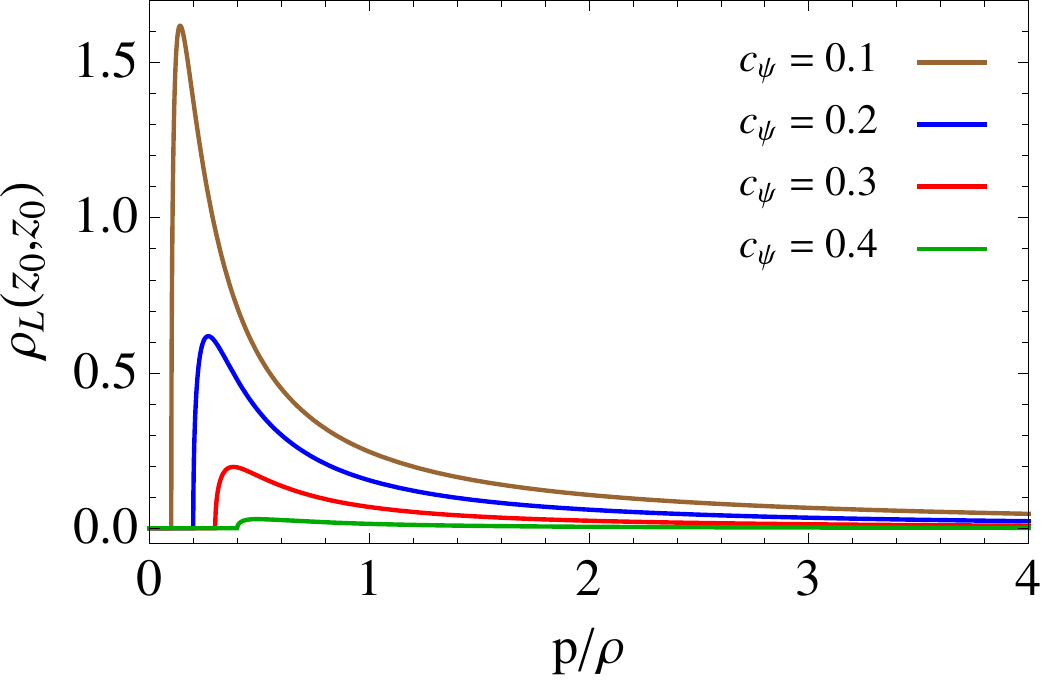} &
\includegraphics[width=0.43\textwidth]{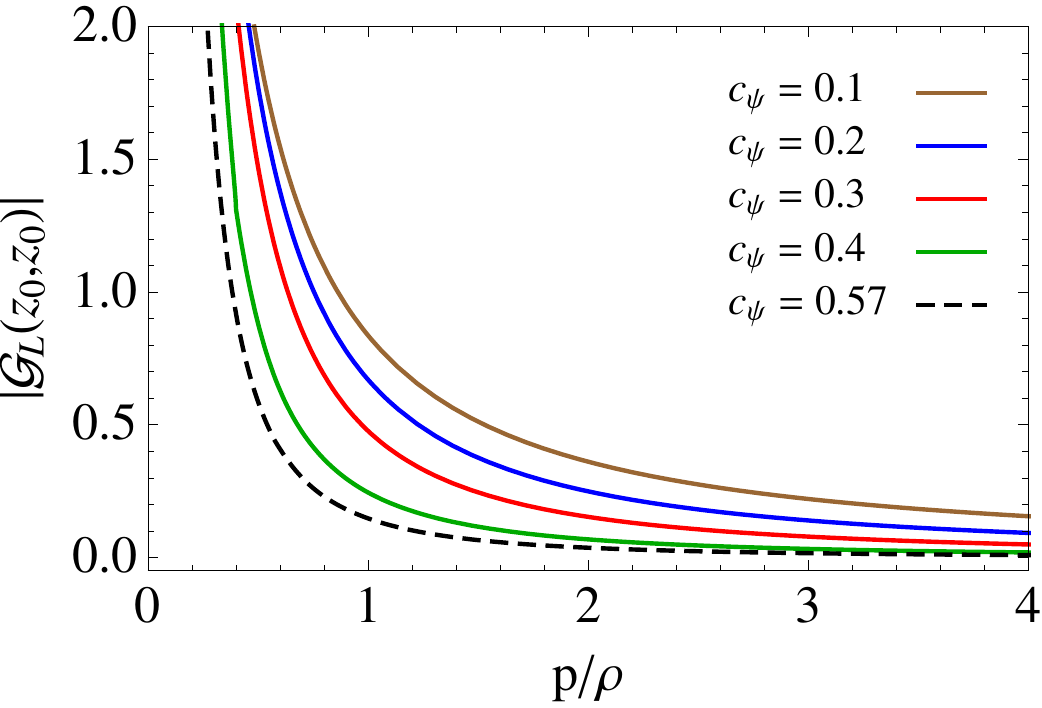} 
\end{tabular}
 \caption{\it Scale invariant spectral density $\rho_L(z_0,z_0;p)$ (left panel) and absolute value of the Green's function $\mathcal G_L(z_0,z_0;p)$ (right panel) for a continuum massless left-handed fermion. We have used $c_\psi = 0.1, 0.2, 0.3, 0.4$ (solid lines) and $0.57$ (dashed line).}
\label{fig:fermions}
\end{figure*}

In the case of fermions~$\psi=(\psi_L,\psi_R)^T$, the Lagrangian is 
\begin{equation}
\mathcal L=\int_0^{y_s}dy \left[e^A\, i\bar\psi /\hspace{-.23cm}\partial\psi-M_\psi(y)\bar\psi\psi + (\bar\psi_R\psi^\prime_L+\bar\psi_L^\prime\psi_R) \right] \,,
\end{equation}
where $M_\psi=\mp \kappa^2 W/6$,~\footnote{Where the sign $-$ ($+$) corresponds to fermions with left handed (right handed) zero modes.} is a bulk mass. The EoM of the fluctuations can be written in the form
\begin{equation}
-\ddt{\psi}_{L,R}(z) + V_{L,R}(z) {\psi}_{L,R}(z) = p^2 {\psi}_{L,R}(z)  \,,  \quad V_{L,R}(z)   \stackrel[z \to \infty]{\longrightarrow}{}  \left( c_\psi \rho \right)^2 \,, \label{eq:EOM_psi}
\end{equation}
and the effective Schr\"odinger potential $V_{L,R}$ turns out to have a mass gap given by $m_\rho = c_\psi \rho$. Following a similar procedure to the one presented in Sec.~\ref{subsec:gauge_bosons} for gauge bosons, one can define the fluctuations in momentum space~$\psi_{L,R}(p,z)=f_{L,R}(p,z) \psi^{(4)}_{L,R}(p)$, and the holographic Lagrangian turns out to be~\footnote{We use the notation $\sigma^\mu = (1,\sigma^i)$ and $\bar\sigma^\mu = (1,-\sigma^i)$, with $\sigma^i \; (i=1,2,3)$ the Pauli matrices.}
\begin{equation}
\mathcal L_{\hol}=-\frac{f_R(p,z_0)}{f_L(p,z_0)}\frac{\bar\psi_L^0\bar\sigma^\mu p_\mu \psi^0_L}{p} \,, \label{eq:Lhol_fermions}
\end{equation}
with $\psi_L(p,z_0)\equiv \psi_L^0(p)$ playing the role of a left-handed source coupled to the right-handed CFT operator $\mathcal O_R$, i.e.~$\bar\psi^0_L\mathcal O_R+\bar{\mathcal O}_R\psi^0_L$.~\footnote{The same analysis can be done for a right-handed source $\psi_R^0(p)$ coupled to a left-handed CFT operator $\mathcal O_L$. In this case, the results are the same as in Eqs.~(\ref{eq:Lhol_fermions}) and (\ref{eq:GL}) but replacing $f_L \to \bar f_R$ and $f_R \to \bar f_L$.} The two-point function is then $S_L=\sigma^\mu p_\mu\, G_L(z_0,z_0;p)$ with the 4D Green's function and spectral density given by
\begin{equation}
G_L(z_0,z_0;p)=-\frac{1}{p} \frac{f_L(p,z_0)}{f_R(p,z_0)} \,, \qquad \rho_L(z_0,z_0;p)=\frac{1}{\pi}\, \textrm{Im} \, G_L(z_0,z_0;p) \,.  \label{eq:GL}
\end{equation}
The scale invariant Green's function, $\mathcal G_{L}(z_0,z_0;p)\equiv \rho^{1+2 c_\psi}G_{L}(z_0,z_0;p)$, and spectral density are displayed in Fig.~\ref{fig:fermions} for different values of $c_\psi$. A comparison with the propagator of unparticles $\mathcal O_{L}$ with dim~$d$, i.e.~$\Delta_L\propto (-p^2-i\epsilon)^{d-5/2}\sigma^\mu p_\mu$ translates into the dimension for the operator $\mathcal O_{L}$, $d_{L}=2 - c_\psi$, which is in agreement with the general results~\cite{Cacciapaglia:2008ns}.

\subsection{The graviton and radion}
\label{subsec:graviton}

The graviton is a transverse traceless fluctuation of the metric of the form
\begin{equation}
ds^2 = e^{-2A(y)} (\eta_{\mu\nu} + h_{\mu\nu}(x,y)) dx^\mu dx^\nu - dy^2 \,.
\end{equation}
By using the ansatz $h_{\mu\nu}(x,y) =\h(y) h_{\mu\nu}(x)$, the EoM for the fluctuation can be written in a Schr\"odinger like form similar to Eqs.~(\ref{eq:EOM_fA}) and (\ref{eq:EOM_psi}). Then, the scale invariant 4D Green's function and spectral density turn out to be
\begin{equation}
\mathcal G_\h (z_0,z_0;p)= \frac{\rho^2}{k} \frac{\h(p,z_0)}{\dt{\h}(p,z_0)} \,, \qquad \rho_\h(z_0,z_0;p)=\frac{1}{\pi}\textrm{Im} \,  \mathcal G_\h(z_0,z_0;p) \,. \label{eq:Ggraviton}
\end{equation}

\begin{figure*}[htb]
\centering
 \begin{tabular}{c@{\hspace{3.5em}}c}
 \includegraphics[width=0.43\textwidth]{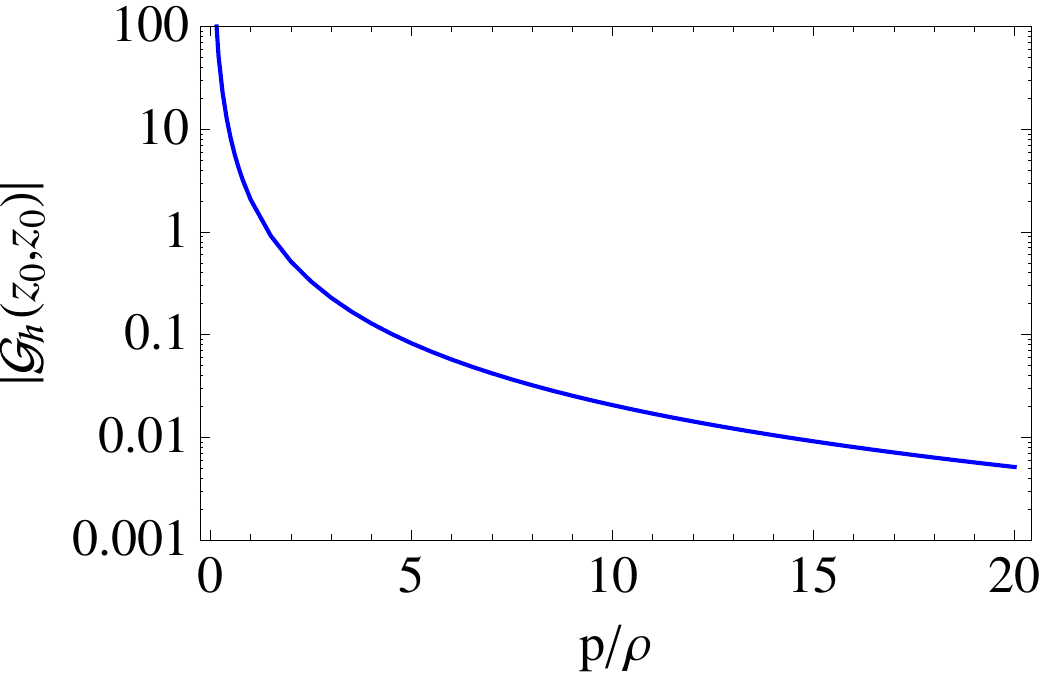} &
\includegraphics[width=0.43\textwidth]{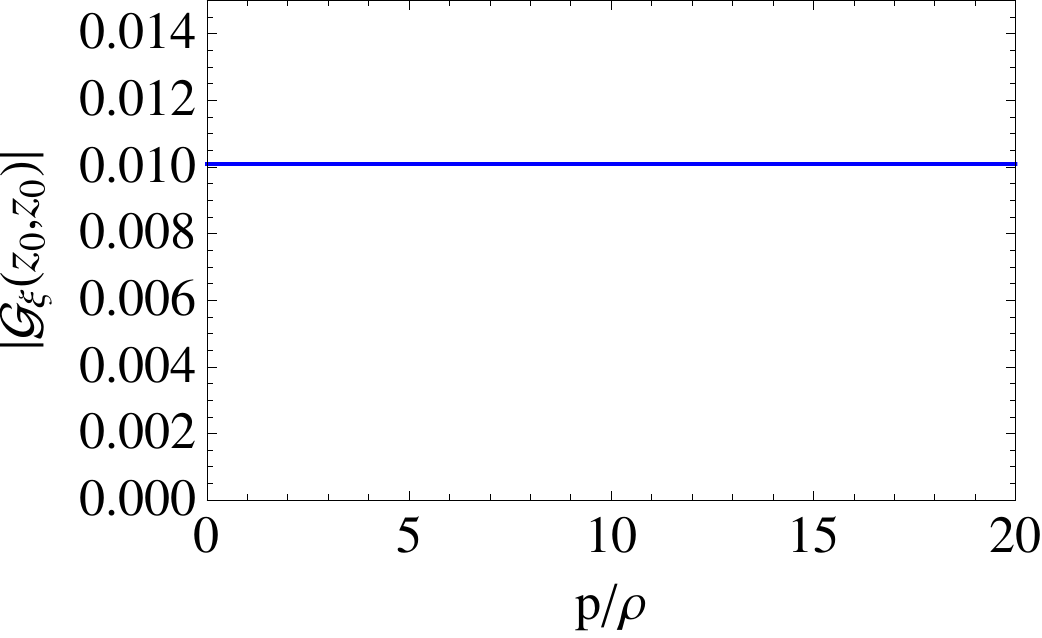} 
\end{tabular}
 \caption{\it Absolute value of the Green's function for a continuum graviton (left panel) and for a continuum radion (right panel), as given by Eqs.~(\ref{eq:Ggraviton}) and (\ref{eq:Gradion}), respectively.}
\label{fig:graviton_radion}
\end{figure*}

Regarding the radion field $\xi(x,y)$, it is defined as the scalar perturbation~\cite{Csaki:2000zn}
\begin{equation}
\hspace{-2.0cm} \phi(x,y)=\phi(y)+\delta\phi \quad \textrm{and}\quad ds^2=-N^2dy^2+g_{\mu\nu}(dx^\mu+N^\mu dy)(dx^\nu+N^\nu dy) \,,
\end{equation}
where~$g_{\mu\nu}=e^{-2A-2\xi}\,\eta_{\mu\nu}$. Using the holographic procedure, the scale invariant Green's function at momenta $p \sim {\cal O}(\rho)$ turns out to be
\begin{equation}
\mathcal G_\xi^{-1}(z_0,z_0;p)  \simeq  \frac{3}{2} \mathcal W^{-2}(k/\rho)  (\rho/k)^{2}  (p/\rho)^2 - 8 k^{-1} \bar\lambda_0(v_0) \,,  \label{eq:Gradion}
\end{equation} 
where $\bar\lambda_0 \equiv 2\kappa^2 \lambda_0$. This leads to a constant behavior, as the term $\bar\lambda_0$ is by far the dominant contribution at these scales. Only when $p \sim {\cal O}(k)$, the Green's function has a non-constant behavior and in particular it goes to zero when $p\to \infty$. We display in Fig.~\ref{fig:graviton_radion} the scale invariant Green's function for the graviton (left panel) and the radion (right panel). In both cases the spectra present a mass gap~$m_g = (3/2) \rho$.

\subsection{The Higgs boson}
\label{suibsec:Higgs}

\begin{figure*}[t]
 \includegraphics[width=0.43\textwidth]{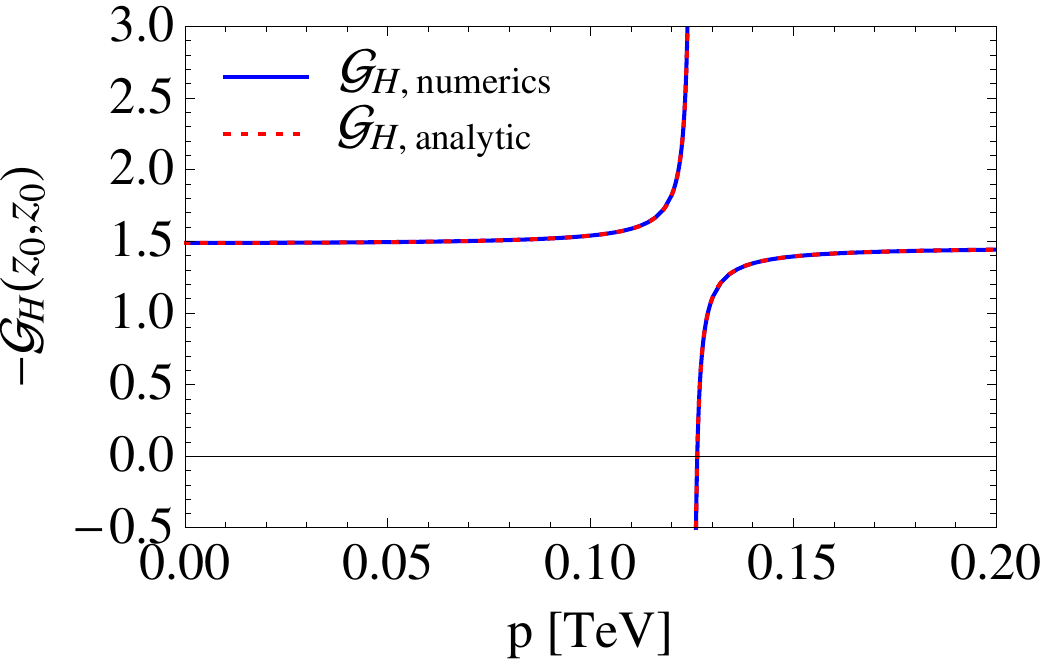} 
\hspace{-1cm} \begin{minipage}{23pc}
\vspace{-4.3cm} \caption{\it Green's function, $\mathcal G_H(z_0,z_0;p)$, for a continuum Higgs boson. We display in solid (blue) line the numerical computation using Eq.~(\ref{eq:GH}), and in dotted (red) line the analytical result from Eq.~(\ref{eq:GHanalytic}). We have used $\rho = 2$ TeV. The pole is located at $p = 0.125 \,\textrm{TeV}$.}
\label{fig:Higgs_boson}
\end{minipage}
\end{figure*}

The action for the physical Higgs boson~$H(x,y) = h(y) + \widehat H(x,y)$ is given by Eq.~(\ref{eq:S_Higgs}). After an appropriate rescaling of the field, one obtains a Schr\"odinger like form for the EoM of the excitation, $\widehat H(x,y)$, with boundary condition in the UV and jump condition in the IR. Using the holographic procedure, the scale invariant Green's function for the Higgs turns out to be
\begin{equation}
\mathcal G_H(z_0,z_0;p) = \left( \frac{1}{k }\frac{\dt{\mathcal H}(p,z_0)}{\mathcal H(p,z_0)} - \frac{2M_0}{k} \right)^{-1}\,. \label{eq:GH} 
\end{equation}
We display in Fig.~\ref{fig:Higgs_boson} the function $\mathcal G_H(z_0,z_0;p)$ in the regime of momenta around the Higgs mass. One can see that the Green's function shows a pole behavior
\begin{equation}
\mathcal G_H(z_0,z_0;p) \simeq c_1 + c_2 \frac{m_H^2}{p^2-m_H^2} \,, \label{eq:GHanalytic}
\end{equation}
with $c_1 = \mathcal G_H(z_0,z_0; p \to \infty)$ and $c_2 =  \mathcal G_H(z_0,z_0;p \to \infty) -  \mathcal G_H(z_0,z_0;p \to 0)  \propto  \rho^2/m_H^2$. This pole corresponds to the presence of an isolated narrow resonance which is identified with the SM Higgs. In addition, there is a continuum of states separated from the resonance by a TeV mass gap, $m_g = (3/2)\rho$.

\section{Phenomenological aspects}
\label{sec:phenomenology}

The aim of this section is to study the consequences of the possible existence of a continuum KK spectrum beyond a mass gap $m_g$, in the model defined by Eq.~(\ref{eq:W_softwall}), for the LHC phenomenology. In particular, the presence of a continuum should be associated with an excess, with respect to the SM prediction, in the cross section of some processes. We summarize in Table~\ref{table} the different mass gaps for the different fields as computed in Sec.~\ref{sec:Green_functions}.
\begin{table}[ht]
\centering
\begin{tabular}{||c||c|c|c|c|c||}
\hline\hline
Field & Gauge boson & Fermion $f$& Graviton & Radion & Higgs \\
\hline
$m_g$ & $\frac{1}{2}\rho$ & $|c_f|\rho$ & $\frac{3}{2}\rho$ & $\frac{3}{2}\rho$& $\frac{3}{2}\rho$\\
\hline\hline
\end{tabular}
\caption{\it Values of the mass gap for different fields, where $\rho\equiv e^{-ky_s}/y_s \sim {\cal O}(\textrm{TeV})$.}
\label{table}
\end{table}
Note that the larger the mass gap, the higher energy should one produce to detect the excess. As for light fermions $c_f > 1/2$, and so their mass gap is $m_g > \rho/2$, it turns out that the simplest case for producing the continuum of KK modes are gauge bosons and, in particular the strongest coupled KK modes, the KK gluons. In the following we will concentrate ourselves in this case.

\subsection{Brane-to-brane Green's function}
\label{subsec:brane_to_brane_Green}

The computation of Green's functions of several fields $G(z_0,z_0;p)$, by using the holographic method, has been presented in Sec.~\ref{sec:Green_functions}. These Green's functions represent the UV-brane-to-UV-brane propagators. There is an alternative procedure to compute more general Green's functions $G(z,z^\prime;p)$ which, in the case of gauge bosons, is based on the inhomogeneous version of Eq.~(\ref{eq:EOM_fA}), i.e.
\begin{equation}
e^{-A} p^2 G_A(z,z';p) + \frac{d}{dz}\left[ e^{-A} \dt{G}_A(z,z';p) \right] = \delta(z-z^\prime) \,, \label{eq:EOM_GA}
\end{equation}
where the dot indicates derivative with respect to~$z$. This equation accounts for the generalization of the Green's function to the continuous spectrum. For the case of a discrete spectrum with mass $m_n$ for the $n$-th mode with wave function $f_A^n$, the Green's function is well-known to be given by
\begin{equation}
G_A(z,z';p)= \sum_n \frac{f_A^n(z)f_A^n(z^\prime)}{p^2-m_n^2} \,, \label{eq:GA_fAn}
\end{equation}
where $\{f_A^n(z)\}$ is a basis of orthonormal modes. After fixing the value of $z^\prime$, the $z$ space is divided into the following domains: $z_0 \le z \le z^\prime$, $z^\prime \le z \le z_1$ and $z_1 \le z < \infty$. The Green's function fulfills a Neumann boundary condition in the UV brane, i.e.~$\dt{G}_A(z_0) = 0$, and is continuous at $z = z_0$, $z^\prime$ and $z_1$. In addition, the inhomogeneous term of the right-hand side of Eq.~(\ref{eq:EOM_GA}) leads to a jump in the derivative, i.e.~$\Delta \dt{G}_A(z^\prime) = e^{A(z^\prime)}$. Finally, we should impose regularity in the IR as explained below Eq.~(\ref{eq:fA_IR}), i.e.~$c_+ = 0$. 

Once the Green's function $G_A(z,z^\prime;p)$ is computed, we can focus on the following three interesting cases: i) the UV-brane-to-UV-brane Green's function 
\begin{equation}
G_A(z_0,z_0;p) = \lim_{z,z^\prime \to z_0} G_A(z,z^\prime; p) \,,
\end{equation}
which coincides with holographic Green's function computed in Sec.~\ref{subsec:gauge_bosons}; ii) the UV-brane-to-IR-brane, and iii) the IR-brane-to-IR-brane Green's functions
\begin{equation}
\hspace{-1.5cm} G_A(z_0,z_1;p) = \lim_{z \to z_0, \, z^\prime \to z_1} G_A(z,z^\prime; p) \,, \qquad G_A(z_1,z_1;p) = \lim_{z, z^\prime \to z_1} G_A(z,z^\prime; p) \,,
\end{equation}
respectively. The scale invariant Green's functions for these cases are $\mathcal G_A(z_{0,1},z_{0,1};p) \equiv$  $(\rho^2/k) \mathcal W(k/\rho) G_A (z_{0,1},z_{0,1};p)$.

\subsection{Enhanced cross sections of Drell-Yan processes at the LHC}
\label{subsec:UVIR_Green}

\begin{figure*}[t]
\centering
 \begin{tabular}{c@{\hspace{3.5em}}c}
 \includegraphics[width=0.43\textwidth]{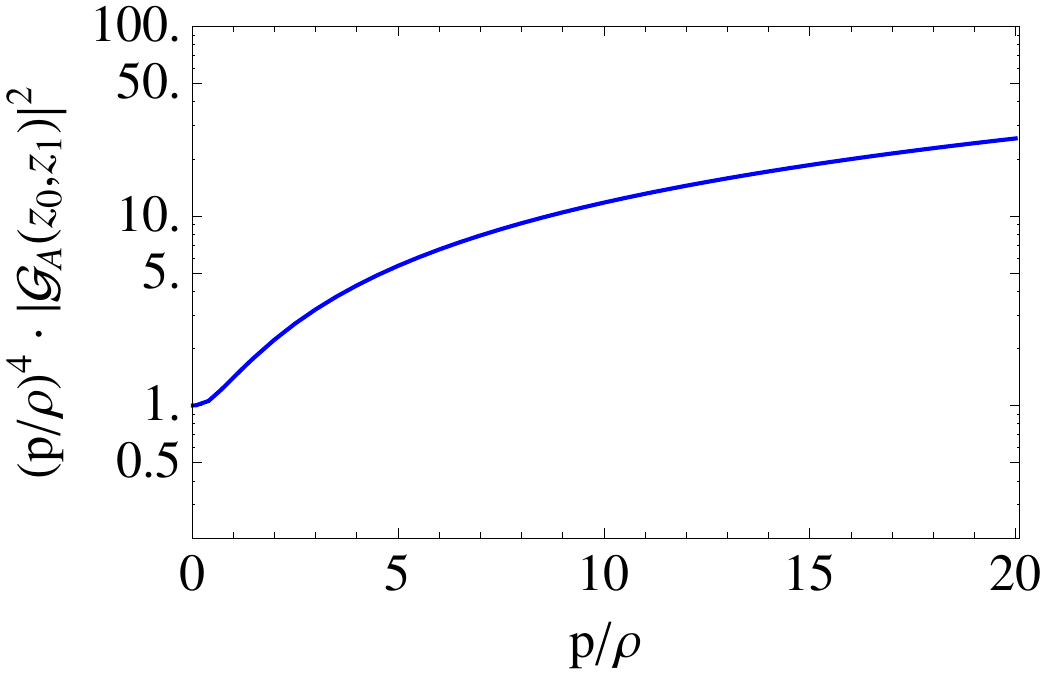} &
\includegraphics[width=0.43\textwidth]{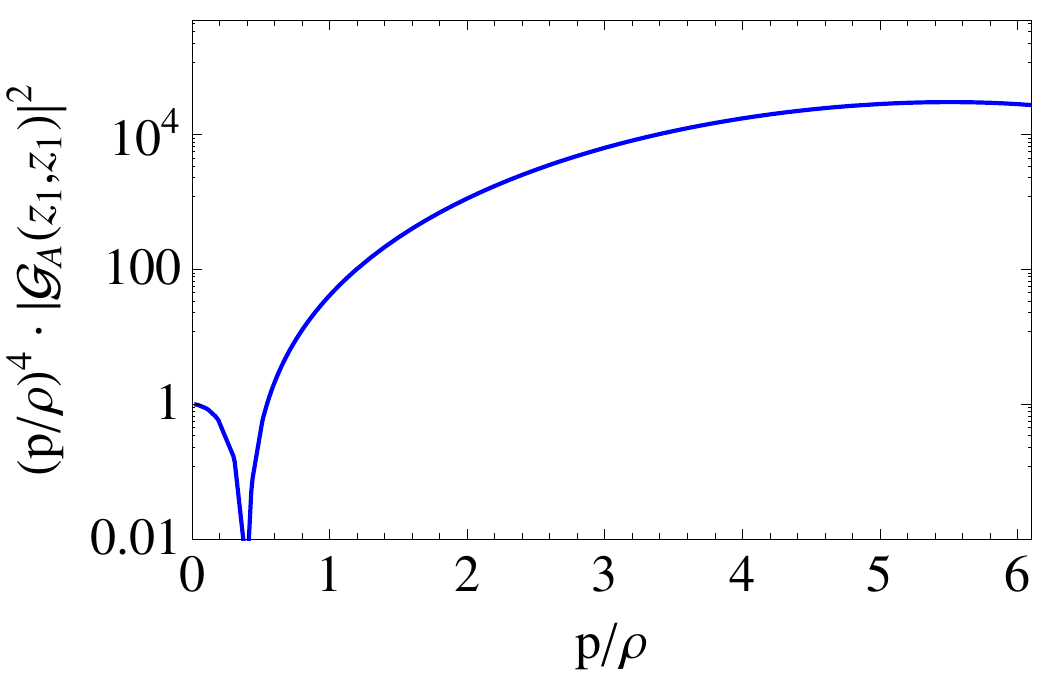} 
\end{tabular}
 \caption{\it Left  panel: $\sigma(q\bar q\to g^*\to f_{\IR}\bar f_{\IR})/\sigma_{\SM}(q\bar q\to g^0\to f_{\IR} \bar f_{\IR})$, versus $p/\rho$, where $f_{\IR}$ is a heavy quark living on the IR brane. Right panel: $\sigma( b_R\bar b_R \to g^*\to t_R \bar t_R) / \sigma_{\SM}( b_R\bar b_R \to g^{(0)}\to  t_R\bar t_R)$, versus $p/\rho$. $p\equiv\sqrt{\hat s}$ is the partonic energy.}
\label{fig:enhancement}
\end{figure*}

In Drell-Yan (DY) processes the continuum of KK gluons is produced by pairs of light fermions $(q \bar q)$, which we can assume to be localized on the UV brane at $z=z_0$. Subsequently, the continuum will decay into a pair of light/heavy fermions $(f_{\UV}\bar f_{\UV})/(f_{\IR}\bar f_{\IR})$ which, as indicated, are localized in the UV/IR brane respectively. The UV-brane-to-UV/IR-brane Green's function $\mathcal G_A(z_0,z_{0,1};p^2)$ can be used to compute the contribution of the gauge continuum $g^*$ to this process. In particular, the excess in the cross section with respect to the SM prediction is given by
\begin{eqnarray}
\hspace{-2cm} &&\sigma(q\bar q \to g^*\to f_{\UV}\bar f_{\UV}) / \sigma_{\SM}( q\bar q \to g^{(0)}\to f_{\UV}\bar f_{\UV} ) = |(\hat s/\rho^2) \, \mathcal G_A(z_0,z_0;\hat s)|^2  \,,  \label{eq:cross_sections_qUV} \\
\hspace{-2cm} &&\sigma( q\bar q \to g^*\to f_{\IR}\bar f_{\IR}) / \sigma_{\SM}( q\bar q \to g^{(0)}\to f_{\IR}\bar f_{\IR} ) =  |(\hat s/\rho^2) \, \mathcal G_A(z_0,z_1;\hat s)|^2 \,, \label{eq:cross_sections_qIR}
\end{eqnarray}
where $p = \sqrt{\hat s}$ is the partonic center of mass energy, $g^*$ is the contribution from the gluon continuum, and $g^{(0)}$ is the SM gluon. $q$ is a proton valence quark living on the UV brane, and $f_{\UV/\IR}$ is either a light quark (as e.g. $b_{L,R}$) or a heavy quark (as e.g. $t_R$). We show in the left panel of Fig.~\ref{fig:enhancement} the case where $f_{\IR} = t_R$. As fermions localized on the IR brane are strongly coupled to the KK modes, the relative cross section with respect to the SM one increases with the partonic energy, yielding an enhancement that can be~$\mathcal O(10)$ for  $p\simeq \mathcal O(10)\rho$. The enhancement with respect to the SM prediction in the case $f_{\UV} = b_{L,R}$ turns out to be $\mathcal O(1)$, and thus it is much more difficult to detect experimentally.

We will also consider the case of the IR-brane-to-IR-brane Green's function $\mathcal G_A(z_1,z_1;p^2)$, which is relevant in processes where both the initial and final fermions are localized on the IR brane. This is the case for instance in models explaining the $R_{D^{(\ast)}}$ anomalies~\cite{Megias:2017ove,Carena:2018cow}. This Green's function can contribute significantly to the process
\begin{equation}
\hspace{-2cm} \sigma( b_R\bar b_R \to g^*\to t_R \bar t_R) / \sigma_{\SM}( b_R\bar b_R \to g^{(0)}\to  t_R\bar t_R) =  |(\hat s/\rho^2) \, \mathcal G_A(z_1,z_1;\hat s)|^2 \,, \label{eq:cross_sections_bR}
\end{equation}
due to the large coupling of $b_R$ to the KK gluon modes in those models. We display this ratio in the right panel of Fig.~\ref{fig:enhancement}. One can see that the enhancement of the production through the gluon continuum can easily be $\mathcal O(10^2-10^4)$. Although this process is parton distribution function suppressed with respect to $\sigma(q \bar q\to g^*\to t_R\bar t_R)$ by the small amount of bottoms inside the proton, the effect of the continuum KK gluon can lead to a strong deviation with respect to the SM predictions for large collider energies. Finally, notice that all these processes, Eqs.~(\ref{eq:cross_sections_qUV})-(\ref{eq:cross_sections_bR}), are dominated  in the limit $p \to 0$ by the gluon zero mode, so that there is no enhancement at these energies as it can be seen from Fig.~\ref{fig:enhancement}.

\section{Continuum linear dilaton model: Green's functions and effective theory}
\label{sec:Linear_Dilaton_Green}

In this section we will provide analytical results for the Green's function of gauge bosons in the LD model of Sec.~\ref{subsec:linear_dilaton}, and discuss the low energy effective theory.

\subsection{Green's functions for massless gauge bosons}
\label{subsec:gauge_bosons_LinearDilaton}

In the LD model the Green's function for gauge bosons obeys the EoM of Eq.~(\ref{eq:EOM_GA}), which can be solved analytically. The general solution of the homogeneous part is
\begin{equation}
\hspace{-2cm} G_{A,\LD}(z,z^\prime;p) =  \left\{ 
\begin{array}{cc}
C_1^I e^{\frac{1}{2}\left(1 - \Delta \right) \rho \cdot (z-z_0)} + C_2^I e^{\frac{1}{2}\left(1 + \Delta \right) \rho \cdot (z-z_0)}    & \quad z_0 \le z  \le   z^\prime  \\
C_1^{II} e^{\frac{1}{2}\left(1 - \Delta \right) \rho \cdot (z-z_0)} + C_2^{II} e^{\frac{1}{2}\left(1 + \Delta \right) \rho \cdot (z-z_0)}  & \quad  z^\prime < z < \infty 
\end{array} \,, \right. \label{eq:GALD_gen}
\end{equation}
where $\Delta$ is defined in Eq.~(\ref{eq:fA_IR}) and $\rho=1/y_s$. The Green's function is subject to the following boundary and matching conditions
\begin{equation}
\hspace{-2cm} \dt{G}_{A,\LD}(z_0,z') = 0 \,, \qquad  \Delta G_{A,\LD}(z^\prime,z') = 0 \,, \qquad  \Delta \dt{G}_{A,\LD}(z^\prime,z') = e^{A(z^\prime)} \,,
\end{equation}
in addition to regularity in the IR. Using these conditions one finds~\footnote{Note that we could have split the domain $z^\prime < z < \infty$ into two domains: $z^\prime < z \le z_1$ and $z_1 < z < \infty$. However, in that case  the boundary conditions in the IR brane, $z = z_1$, i.e.~continuity of $G_{A,\LD}(z,z')$ and $\dt{G}_{A,\LD}(z,z')$, demand that the term $\propto e^{\frac{1}{2} \left( 1 + \Delta\right) \rho \cdot (z - z_0)}$ is also absent in  $z^\prime < z \le z_1$, so that the solution would be identical as the one presented in Eq.~(\ref{eq:GALD_gen}) with constants given by Eq.~(\ref{eq:Cs}).} 
\begin{eqnarray}
\hspace{-2cm} C_1^{I} &=& \frac{(1+\Delta)}{(1-\Delta)\Delta \cdot \rho}  e^{\frac{1}{2}\left( 1 - \Delta \right) \rho \cdot (z^\prime - z_0)} \,, \qquad \hspace{2.5cm} C_2^{I} = -\frac{1}{\Delta \cdot \rho}  e^{\frac{1}{2}\left( 1 - \Delta \right) \rho \cdot (z^\prime - z_0)} \,, \nonumber \\
\hspace{-2cm} C_1^{II} &=&  \frac{1}{\Delta\cdot \rho} \left( \frac{1+\Delta}{1-\Delta} e^{\frac{1}{2}(1-\Delta) \rho \cdot (z^\prime - z_0)} - e^{\frac{1}{2}(1+\Delta) \rho \cdot (z^\prime -z_0)} \right) \,, \quad C_2^{II} = 0  \,. \label{eq:Cs}
\end{eqnarray}
The Green's function in proper coordinates writes as in Eqs.~(\ref{eq:GALD_gen}) and (\ref{eq:Cs}) after replacing $e^{\rho \cdot (z^{(\prime)} - z_0)} \to (1-y^{(\prime)}/y_s)^{-1}$. One can write the following analytical expressions for the relevant UV-brane-to-UV/IR-brane Green's functions studied in Sec.~\ref{sec:phenomenology},
\begin{eqnarray}
G_{A,\LD}(z_0,z_0;p) &=&  \frac{2}{(1-\Delta)\rho}  \,, \label{eq:GALD_z0z0}   \\
G_{A,\LD}(z_0,z_1;p) &=&  \frac{2}{(1-\Delta)\rho} e^{\frac{1}{2}(1-\Delta) A_1}  \,,   \label{eq:GALD_z0z1}  \\
G_{A,\LD}(z_1,z_1;p) &=& \frac{1}{\Delta \cdot \rho} \left( \frac{1+\Delta}{1-\Delta} e^{(1-\Delta) A_1}   - e^{A_1}  \right) \label{eq:GALD_z1z1}  \,.
\end{eqnarray}
The behavior of these Green's functions in the regime $p \ll \rho$ is
\begin{eqnarray}
G_{A,\LD}(z_0,z_0;p)^{-1} &\stackrel[p \ll \rho]{\simeq}{}&  \frac{p^2}{\rho} + \frac{p^4}{\rho^3} + {\cal O}(p^6) \,, \label{eq:GALD_z0z0_psmall}   \\
G_{A,\LD}(z_0,z_1;p)^{-1} &\stackrel[p \ll \rho]{\simeq}{}&  \frac{p^2}{\rho} + (1 - A_1)  \frac{p^4}{\rho^3} + {\cal O}(p^6) \,,   \label{eq:GALD_z0z1_psmall}  \\
G_{A,\LD}(z_1,z_1;p)^{-1} &\stackrel[p \ll \rho]{\simeq}{}&  \frac{p^2}{\rho} + \left( e^{A_1} - 2 A_1 \right) \frac{p^4}{\rho^3} + {\cal O}(p^6) \,. \label{eq:GALD_z1z1_psmall} 
\end{eqnarray}
We have used in these expressions that $A_1 = \rho \cdot (z_1 - z_0)$, cf.~Eq.~(\ref{eq:phiALD}). The Green's functions have a pole behavior a low momentum, $G_{A,\LD}(p) \sim 1/p^2$, as it is expected by the existence of a zero mode. The scale invariant Green's functions for these cases are~$\mathcal G_{A,\LD}(z_{0,1},z_{0,1};p) \equiv \rho \, G_{A,\LD}(z_{0,1},z_{0,1};p)$.

\begin{figure*}[t]
 \includegraphics[width=0.43\textwidth]{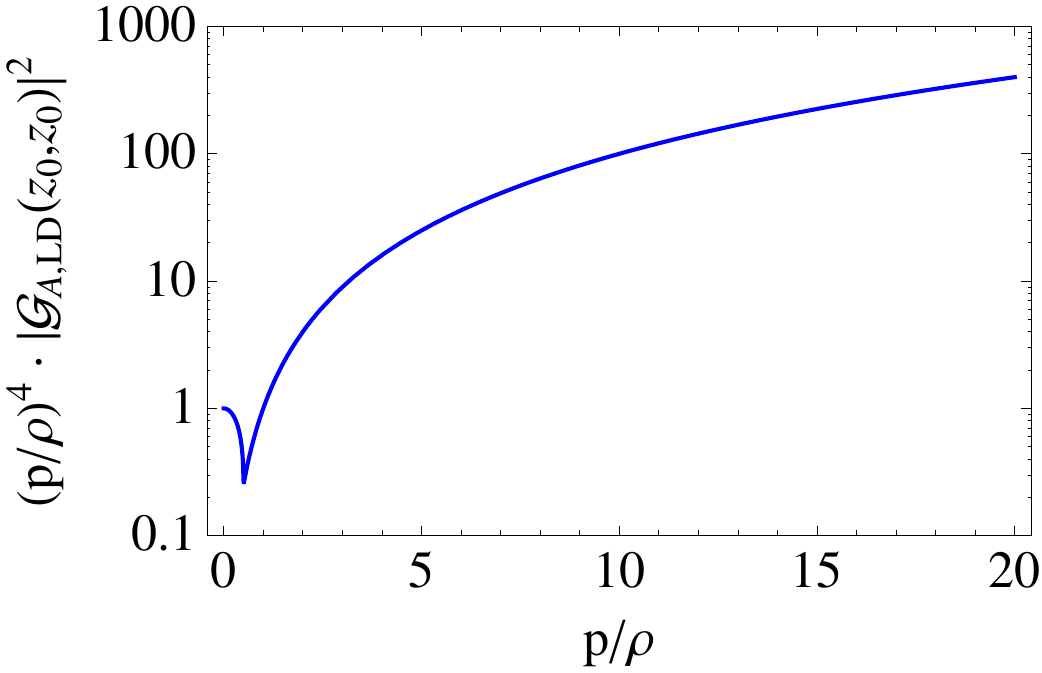} 
\hspace{-1cm} \begin{minipage}{23pc}
\vspace{-3.3cm} \caption{\it  $\sigma(q\bar q\to g^*\to f_{\UV}\bar f_{\UV})/\sigma_{\SM}(q\bar q\to g^0\to f_{\UV} \bar f_{\UV})$  for the LD model as given by Eq.~(\ref{eq:GALD_z0z0}). See Fig.~\ref{fig:enhancement} for further details. }
\label{fig:enhancementLD}
\end{minipage}
\end{figure*}

Using the results of Eqs.~(\ref{eq:GALD_z0z0})-(\ref{eq:GALD_z1z1}), we can see that the Green's functions $\mathcal G_{A,\LD}(z_{0},z_{1};p)$ and $\mathcal G_{A,\LD}(z_{1},z_{1};p)$ have strong enhancements for $p\gg \rho$, which are $\propto e^{A_1}$, compared with the SM prediction by the zero-mode gluon exchange, typically much larger than the ones obtained in the soft wall model presented in Sec.~\ref{sec:phenomenology}. Therefore  if we insist in solving the hierarchy problem with the LD model, i.e.~$A_1 \simeq 35$, then there should be a huge  enhancement of $\sigma(q\bar q\to g^*\to f_{\IR}\bar f_{\IR})$ and $\sigma( b_R\bar b_R \to g^*\to t_R \bar t_R)$ with respect to the corresponding SM cross sections, making the model phenomenologically unappealing. We then conclude that, in the model presented in this section, fermions should be all localized toward the UV brane, what was called $f_{\UV}$, and we should give up explaining the fermion mass hierarchy with different fermion localizations in the warped dimension. We plot in Fig.~\ref{fig:enhancementLD} the ratio of cross sections for the DY process with respect to the SM expectation $\sigma(q\bar q\to g^*\to f_{\UV}\bar f_{\UV})/\sigma_{\SM}(q\bar q\to g^0\to f_{\UV} \bar f_{\UV})$. We see a moderate enhancement that can be~$\mathcal O(10^2)$ for  $p\simeq \mathcal O(10)\rho$ which should be useful to contrast the present model with experimental data. 

\subsection{Low energy effective theory for gauge bosons}
\label{sec:Low_energy_gauge_bosons}

We can also provide some insights into the low energy effective theory obtained after integrating out the KK gauge boson modes. Let us consider the diagram corresponding to the exchange of KK gauge bosons between two pairs of quarks, which we will denote by $q\bar q$ and $t \bar t$. By integrating all the modes except the zero mode, the low energy effective theory turns out to be given by the effective operator
\begin{equation}
\mathcal L_{\EFT}^{(0)} = \frac{C_{qt}^{(0)}(z,z^\prime)}{4\rho^2} \, (\bar q \gamma^\mu \lambda^a q) (\bar t \gamma_\mu\lambda_a t) \,, \label{eq:Wilson}
\end{equation}
where $z$ and $z^\prime$ are the locations of the quarks $q$ and $t$, respectively, and $\lambda^a$ are the eight Gell-Mann matrices. Let us define the function
\begin{equation}
\widehat G_A(z,z^\prime;p) := G_A(z,z^\prime;p) - \frac{\rho}{p^2} \,, \label{eq:G_hat}
\end{equation}
i.e.~the Green's function with the zero mode subtracted out, cf.~Eqs.~(\ref{eq:GALD_z0z0_psmall})-(\ref{eq:GALD_z1z1_psmall}). Then, the Wilson coefficient of the effective Lagrangian Eq.~(\ref{eq:Wilson}) is given by
\begin{equation}
C_{qt}^{(0)}(z,z^\prime) = \lim_{p \to 0} g_5^2 \rho^2 \widehat G_A(z,z^\prime;p) \,, \label{eq:Cqt_general}
\end{equation}
where the 5D $(g_5)$ and 4D $(g_4)$ gauge couplings are correspondingly related by $g_5 = \sqrt{y_1} \, g_4$. When computing the Wilson coefficient within the LD model, one obtains
\begin{equation}
C_{qt}^{(0)}(z_0,z_0) = - g_4^2 \left(1 - e^{-A_1}\right) \,.
\end{equation}
Finally, let us comment that we could also study the effective theory obtained by integrating all the modes above some scale $\Lambda$, i.e.
\begin{equation}
\mathcal L_{\EFT}^\Lambda = \frac{C_{qt}(z_0,z_0;\Lambda)}{4\Lambda^2} \, (\bar q \gamma^\mu \lambda^a q) (\bar t \gamma_\mu \lambda_a t) \,. \label{eq:Wilson_s}
\end{equation}
The corresponding Wilson coefficients can be obtained in a way similar to Eqs.~(\ref{eq:G_hat}) and (\ref{eq:Cqt_general}) after subtracting out not only the zero mode but all the modes below the scale~$\Lambda$ ($m < \Lambda$). We leave this study for a future work~\cite{Megias2020}.

\section{Conclusions}
\label{sec:conclusions}

In this work we have proposed a 5D warped model that solves the hierarchy problem in the conventional fashion, where the 5D Planck scale $M$ and the curvature parameter $k$ are of the order of magnitude of $M_{\rm Planck}$, while the TeV scale is derived from them by the geometrical warp factor. The model leads to a continuum of states with a mass gap for the KK spectra of all the fields (gauge bosons, fermions, graviton, radion and Higgs boson). It predicts the existence of a metric singularity in proper coordinates $y = y_s$, which is admissible as it supports finite temperature. The main feature of the model is that the geometry is AdS$_5$ near the UV brane, and it behaves like the linear dilaton theory near the IR, leading to a strong breaking of conformality in this regime. 

We have computed in this model the holographic Green's functions $G(z_0,z_0;p)$  and spectral densities $\rho(z_0,z_0;p)$ for all the particles. Moreover, as a particular phenomenological application we have studied the generalization to Green's functions $G(z,z';p)$ and in particular the UV/IR-brane-to-UV/IR-brane Green's functions, and seen how they can modify the SM cross section~$\sigma(pp\to Q\bar Q)$ due to the production of a continuum KK gluon by Drell-Yan processes. This leads to an enhancement in the cross section that can be large for momenta $p\gg \rho$. 

Finally we have considered a continuum version of the linear dilaton model, for which the metric and the bulk scalar $\phi$ are exactly linear in conformally flat coordinates. We have studied the dynamical stabilization of the brane distance in this model as well as the solution to the EW hierarchy problem, which can be done without fine-tuning of the parameters. We have computed the general Green's functions and shown that, to avoid huge enhancement with respect to the corresponding SM Green's functions, fermions should be all localized toward the UV brane, thus preventing a solution to the fermion flavor problem by means of different localization of fermions in the extra dimension.  We believe that this is a general feature triggered by the linear behavior of the metric and stabilizing field. 

Other phenomenological applications of these theories should be inspired on unparticle phenomenology~\cite{Georgi:2007si,Cacciapaglia:2007jq,Grinstein:2008qk}. These and other issues will be addressed in a forthcoming publication~\cite{Megias2020}.

%%%%%%%%%%%%%%%%%%%%%%%%%%%
% Acknowledgments
\ack 
%%%%%%%%%%%%%%%%%%%%
The work of EM is supported by the Spanish MINEICO under Grant
FIS2017-85053-C2-1-P, by the FEDER Andaluc\'{\i}a 2014-2020
Operational Programme under Grant A-FQM-178-UGR18, by Junta de
Andaluc\'{\i}a under Grant FQM-225, by Consejer\'{\i}a de
Conocimiento, Investigaci\'on y Universidad of the Junta de
Andaluc\'{\i}a and European Regional Development Fund (ERDF) under
Grant SOMM17/6105/UGR, and by the Spanish Consolider Ingenio 2010
Programme CPAN under Grant CSD2007-00042. The research of EM is also
supported by the Ram\'on y Cajal Program of the Spanish MINEICO under
Grant RYC-2016-20678. The work of MQ is partly supported by Spanish
MINEICO under Grant FPA2017-88915-P, by the Catalan Government under
Grant 2017SGR1069, and by Severo Ochoa Excellence Program of MINEICO
under Grant SEV-2016-0588.
%\vspace*{-1cm}
%\vfill \eject
%%%%%%%%%%%

%%%%%%%%%%%
\section*{References}
%\input{refs}
%\bibliographystyle{elsarticle-num}
%\bibliography{refs}
%%%%%%%%%%%

\end{document}